\documentclass[conference]{IEEEtran}

\IEEEoverridecommandlockouts
\usepackage{cite}
\usepackage{amsmath,amssymb,amsfonts}
\usepackage{algorithmic}
\usepackage{graphicx}
\usepackage{url}
\usepackage{textcomp}
\usepackage{xcolor}
\usepackage[ruled,vlined,nofillcomment,linesnumbered,titlenotnumbered]{algorithm2e}

\usepackage{fancyhdr}
\fancypagestyle{firstpage}{
  \fancyhf{}
  \fancyhead[C]{ 	To appear at the IEEE International Symposium on Circuits and Systems (ISCAS 2025), May 25-28, 2025, London, UK.}
 
}

\usepackage{subcaption,tabularx,booktabs,arydshln}
\newcolumntype{C}{>{\centering\arraybackslash}X}%
\newcolumntype{P}{>{\centering\arraybackslash}X}%
\newcolumntype{R}{>{\raggedleft\arraybackslash}X}%

\usepackage[switch]{lineno}
\usepackage{askmaps, sparklines, tikz, etex}

\makeatletter
\newcommand{\hist}[3][10]{
  \newcount\itmCntr%
  \def\nextitem{\def\nextitem{}}%
  \def\incCntr{\itmCntr=\numexpr\itmCntr+1\relax}%
\begin{sparkline}{#1}
  \@for \el:=#3 \do{\nextitem\pgfmathparse{(\the\itmCntr*1.0+0.5)/#2}\incCntr\let\x=\pgfmathresult\let\y=\el%
      \sparkspike {\x} {\y} 
  }%
\end{sparkline}%
}
\makeatother
\newcommand{\xth}[1]{#1\textsuperscript{th}}
\newcommand*{\Comb}[2]{\binom{#1}{#2}}%
\newcommand{\repository}{\textcolor{blue}{\url{https://github.com/ehw-fit/approximate-medians}}}

\begin{document}

\title{AxMED: Formal Analysis and Automated Design of Approximate Median Filters using BDDs
\thanks{The work on this paper was supported by the Czech Science Foundation grant 24-10990S.}\vspace{-0.5em}
}

\author{\IEEEauthorblockN{Vojtech Mrazek and Zdenek Vasicek}
\IEEEauthorblockA{\textit{Brno University of Technology}, Brno, Czechia \\
mrazek@fit.vutbr.cz, vasicek@fit.vutbr.cz\vspace{-1.5em}}
}

\maketitle

\thispagestyle{firstpage}

\begin{abstract}
The increasing demand for energy-efficient solutions has led to the emergence of an approximate computing paradigm that enables power-efficient implementations in various application areas such as image and data processing. 
The median filter, widely used in image processing and computer vision, is of immense importance in these domains. 
We propose a systematic design methodology for the design of power-efficient median networks suitable for on-chip or FPGA-based implementations. 
A search-based design method is used to obtain approximate medians that show the desired trade-offs between accuracy, power consumption and area on chip.
A new metric tailored to this problem is proposed to quantify the accuracy of approximate medians. Instead of the simple error rate, our method analyses the rank error.
A significant improvement in implementation cost is achieved. For example, compared to the well-optimized high-throughput implementation of the exact 9-input median, a 30\% reduction in area and a 36\% reduction in power consumption was achieved by introducing an error by one position (i.e., allowing the 4th or 6th lowest input to be returned instead of the median).\vspace{-0.5em}
\end{abstract}


\vspace{-0.5em}
\section{Introduction}
The problem of finding the median, or the selection problem in general, is one of the most fundamental computing problems. The median and its variants are widely used in many applications such as sorting, digital signal processing, image processing, and even machine learning \cite{Ferdowsi:2014,Nikolova:2004:Med,Zhang_arxiv.2205.15867}. 
Efficient execution is required for both hardware and software algorithms. To determine the median, two fundamental approaches are the sorting-based method and the histogram-based method~\cite{Adams}.
Hardware implementation offers a variety of energy-saving opportunities. 
In addition to the architecture optimization, the energy efficiency of HW implementations can be further improved by approximate computing techniques -- the errors are introduced to the computation in order to achieve a reduced energy consumption \cite{Mittal:2016}. Various approaches can be used. In this paper, we focus on functional approximation~\cite{bookaxc} which allows to use the standard synthesis flow.

The median filter is an operation that was approximated even before the advent of approximate computing. A notable example of such an approximation is the so-called Median of Medians (MoM) algorithm, which produces an approximate rather than an exact median~\cite{blum}. 
Other approximate approaches, such as those based on separable sorting networks, operate by sharing intermediate results between inputs in stream processing~\cite{Perrot2022, Adams, salvador18}.
The first paper addressing the approximation of 9-input median filters is by Monajati et al.~\cite{Monajati:2015}. In this work, the authors manually approximated 2-bit magnitude comparators, which were then used to construct 8-bit approximate comparators, a crucial component of the CAS operation in HW implementations. The performance median filters built from the 8-bit approximated CAS operations was evaluated using a circuit simulator on a randomly generated set of input data. 
The main limitation of the previous approach is that the quality depends on the data distribution, leading to non-deterministic and hard to predict behavior when used in a real application. In~\cite{vasicek:sekanina:tec}, Vasicek and Sekanina introduced a method for the automatic design of approximate digital circuits using evolutionary algorithms. The objective of the algorithm is to reduce the number of CAS operations of the accurate median network while maintaining the error below a user-defined threshold. Specifically, they reported results for 9-input and 25-input median circuits. Precise CAS are used. The quality of the approximate networks was assessed using a set of training data, and the hardware cost was estimated based on the assumption that the number of CAS operations linearly correlates with hardware cost. This simplification provides a rough estimate for pipelined architectures, accounting for the presence of registers.
Mrazek et al.~\cite{Vasicek2016} focused on the design of approximate median networks suitable for software implementation, enhancing the precision of the quality evaluation process. The quality of candidate approximate circuits was assessed using all permutations generated from a specific monotone sequence, which helped eliminate the bias introduced by randomly generated training vectors. This approach improved the reliability of the evaluation but does not reflect the hardware parameters. 

In this paper, we propose an improved method for design of approximate medians of high quality suitable for real-time applications.
The primary contribution is \textbf{a formal algorithm} for the data-independent and exact evaluation of the quality of approximate medians, which can be efficiently computed using Binary Decision Diagrams (BDDs)\footnote{BDD-based evaluation algorithm and the proposed approximate circuits are available at \repository}.
We \textbf{reduced the evaluation complexity} from $\mathcal{O}(n!)$ to $\mathcal{O}(2^n)$ by extending the zero-one theorem introduced in~\cite{Knuth:ACP} for exact sorting networks. 
In addition to the error rate, our approach can also determine the average and worst-case errors, as well as the error distribution. 
To estimate the hardware cost, we proposed a new \textbf{HW-oriented estimation method} that reflects parameters of the pipelined circuits.
These approaches were applied in the evolutionary-based \textbf{automated approximation}.

\section{Analyzing Approximate Medians}\label{sec:analysis}
To prove that a comparison network $M$ is the exact median network, we can use the so-called {\em zero-one theorem}~\cite{Knuth:ACP} introduced to prove the validity of sorting networks. This theorem states that if a comparison network with $n$ inputs sorts all $2^n$ input sequences of 0's and 1's in non-decreasing order, it will sort any arbitrary sequence of $n$ elements in non-decreasing order. As a consequence of that, it is sufficient to restrict the inputs to Boolean values. This allows the verification task to be formulated as a pseudo-Boolean Constraint Satisfaction Problem (CSP).
Let $M(x_1,\cdots,x_{n})$ be a comparison network with $n$-inputs ($n$ is always odd in this paper). Let $x_i \in B$ be a Boolean variable, $B = \{0,1\}$. Then, each median network must satisfy that 
\begin{equation}
M(x_1,\cdots,x_{n}) = \begin{cases}
         0 & \text{if } x_1 + \cdots + x_n < (n+1)/2 \\
         1 & \text{if } x_1 + \cdots + x_n \ge (n+1)/2 \\
         \end{cases}\label{eq:pbc}
\end{equation}
\noindent holds for all input combinations, i.e. $\forall \{x_1,\cdots,x_n\} \in \mathrm B^n$.

\subsection{Worst-Case Error Analysis}
In the case of an approximate median, an invalid value is returned for some input combination. 
Taking into account the fact that each comparison network performs a selection process, this means that a different value is selected from the sequence of input elements and returned instead of the median.
To model the error introduced by the approximations, we can measure the distance between the rank of the returned element and the rank of the median, i.e., a fixed value equal to $m = (n+1)/2$. For example, the 3\textsuperscript{rd} largest value is returned instead of 5\textsuperscript{th} largest expected for 9-input median, and the rank distance is 2. This can also be applied in the Boolean domain. Let $M({\bf x})=0$ for ${\bf x}=(1,1,0,1,1)$. Let $\widetilde{\bf x}$ denote the sorted sequence $\bf x$, i.e. $\widetilde{\bf x}=(0,1,1,1,1)$. 
In this case, $M$ returns the first lowest element, i.e., $M({\bf x})=\widetilde{x_1}$, instead of the median, i.e., the 3\textsuperscript{rd} lowest element.
Similarly, let $M({\bf x})=1$ for ${\bf x}=(0,1,0,0,0)$. The $\widetilde{\bf x}=(0,0,0,0,1)$ shows that the \xth{5} lowest item was chosen instead of the median. In both cases the distance between the median and the returned value is $d = 2$.

In general, $M$ can have an asymmetric error distribution. Therefore, it may be advantageous to investigate the left ($d^L$) and right ($d^R$) worst-case distance separately.
The worst-case error analysis problem can be formulated using CSP as follows.
Find maximal $d^L \in \{0\ldots m-1\}$ such that $\exists {\bf x} \in B^n : M({\bf x}) = 0 \land x_1 + \cdots + x_n = n - (m - d^L)$. Similarly, find maximal $d^R \in \{0\ldots m-1\}$ such that $\exists {\bf x} \in B^n : M({\bf x}) = 1 \land x_1 + \cdots + x_n = m - d^R$. 
If $d^L=d^R=0$, $M$ is an accurate median network.
Otherwise, $d^L$ ($d^R$) defines the worst case distance between the obtained value and the median that can occur in practice.

\subsection{Error Distribution Analysis}
In most applications, knowing the worst-case error is not enough, as the probability of its occurrence may be negligible.
For such a scenario, we propose a technique to obtain an error distribution that provides information about the probability of occurrence of errors at different distances.

Let $a^L_i=|\{{\bf x} \in B^n: M({\bf x}) = 0 \land x_1 + \cdots + x_n = n - (m - i)\}|$ be the number of input assignments containing exactly $n - (m - i)$ ones for which zero was incorrectly chosen as the output value provided that $0 < i < m$.
Similarly, let $a^R_i=|\{{\bf x} \in B^n: M({\bf x}) = 1 \land x_1 + \cdots + x_n = m - i\}|$ be the number of input assignments containing exactly $n - (m - i)$ zeroes for which 1 was incorrectly chosen as the output value. Note that $a_0 = a^L_0 = a^R_0$ is the number of input assignments for which $M$ provides the expected output value.
The histogram of error distribution $H(M)=(h^L_{m-1},h^L_{m-2},\ldots, h^L_0=h^R_0, \ldots, h^R_{m-2},h^R_{m-1})$ can be determined as 
\begin{equation*}
h^R_i = \begin{cases}

 a^R_i / \Comb{n}{m-i} - a^R_{i+1} / \Comb{n}{m-i-1} & \text{for } 0 \le i < m-1 \\ 
 a^R_i / \Comb{n}{m-i}                               & \text{otherwise} 
 
 \end{cases}
\end{equation*}
\begin{equation*}
h^L_i = \begin{cases}
 a^L_i / \Comb{n}{m-1+i} - a^L_{i+1} / \Comb{n}{m+i} & \text{for } 0 \le i < m-1 \\ 
 a^L_i / \Comb{n}{m-1+i}                             & \text{otherwise} 
 \end{cases}\label{eq:hist}
\end{equation*}
\noindent where $h_i$ is the probability of the error occurence of distance $i$ and binomial coefficient $\Comb{n}{r} = n!/(n-r)!r!$ gives the total number of combinations containing exactly $r$ ones (zeroes).
The calculation was successfully validated according to the exact exhaustive permutation principle \cite{Vasicek2016} for smaller instances.

\begin{figure}[b]\vspace{-15pt}
\centering\includegraphics[width=0.8\columnwidth]{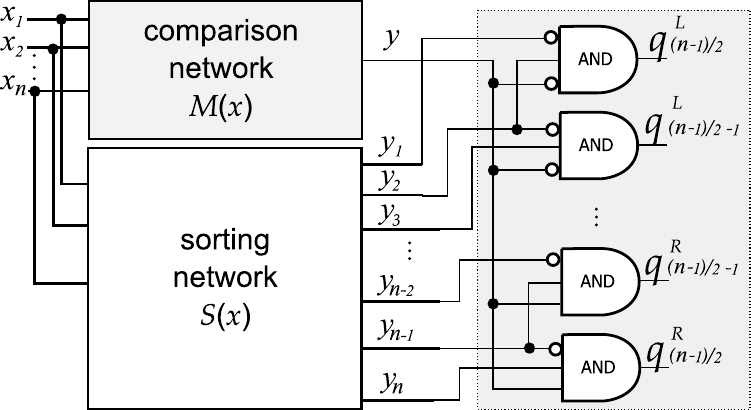}%
\caption{Virtual circuit used for an efficient worst-case and error distribution analysis}%
\label{fig:bdd}\vspace{-1em}
\end{figure} 

\subsection{Analysis using BDDs}

To compute the error distribution and efficiently identify the worst case error, we propose to solve the CSP problem using BDDs and the virtual circuit shown in Figure~\ref{fig:bdd}.
This circuit consists of the analysed comparison network, a sorting network and a simple logic. 
The sorting network acts as a counter of the number of input variables that are set to one. 
To determine $a^L_i$ and $a^R_i$, the SatCount (\#SAT) operation available in every BDD package is called for each output, i.e.
$a^L_i = \mathrm{SatCount}(q^L_i)$, $a^R_i = \mathrm{SatCount}(q^R_i)$, $i=0\ldots m-1$.

The construction of BDD is straightforward. 
Due to the zero-one principle, each CAS element corresponds to a pair of logic gates (AND and OR) that must to be included in the BDD. The AND gate determines the minimum value, the OR gate determines the maximum value. 
Finally, the auxiliary logic is included in the BDD.

\section{Proposed Approximation Method}\label{sec:method}
\begin{figure}[b]
\centering\vspace{-1.5em}
\includegraphics[width=0.9\columnwidth]{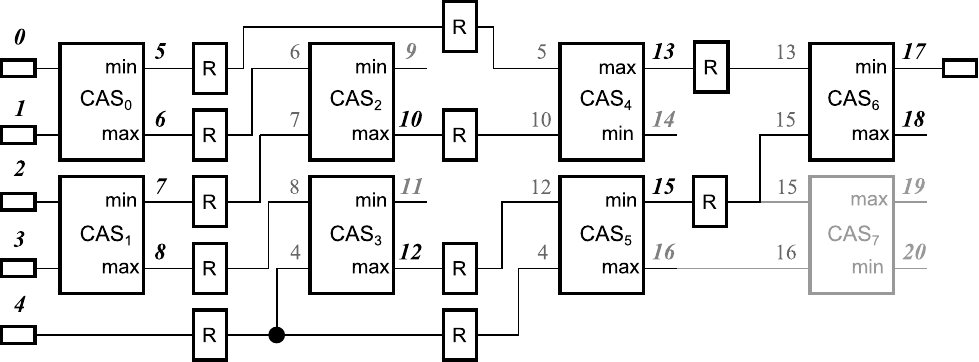}
\caption{Example of a comparison network consisting of 7 operations encoded using CGP with parameters: $n=5$, $n_c=8$.
This network can be encoded using the sequence $(0,1,0)(2,3,0)(6,7,0)(8,4,0)(5,10,1)$ $(12,4,0)(13,15,0)(15,16,1)(17)$. 
}\vspace{-1em}
\label{fig:cgp}
\end{figure}

Several approaches have been proposed to approximate digital circuits~\cite{aproxcomp22, Sekanina2021}. Inspired by~\cite{vasicek:sekanina:tec}, we chose Cartesian Genetic Programming (CGP)~\cite{miller:cgp:book}. 
CGP is an evolutionary algorithm that encodes a circuit to a cartesian grid of nodes. The circuit is represented by an integer netlist specifying the connection of node inputs, functions, and the output node. In our case, a comparison network with $n$ inputs consisting of $k$ CAS elements can be represented by a directed acyclic graph with $k$ two-input two-output nodes, where each node corresponds to a single CAS.
An example is given in Fig.~\ref{fig:cgp}.

The CGP algorithm starts with an exact circuit encoded in the aforementioned encoding. Then, $\lambda$ offsprings are generated using mutation of $h$ randomly chosen integers in the netlist while keeping constraints on generating acyclic representation. A random mutation may disconnect some active nodes that become redundant (see node 7 in the example in Fig.~\ref{fig:cgp}).
For a detailed description, please refer to~\cite {miller:cgp:book}.
We propose to apply a two-stage procedure. At the beginning, the designer specifies the target reduction (area) that should be achieved. The first stage starts with an exact solution. The goal is to gradually modify the initial solution and obtain a reduced network of the target cost $t$, providing that a small difference $\varepsilon$ is tolerated, which prevents the search from getting stuck in a local extreme.
In the second stage, which begins as soon as the target reduction is achieved, the search method reflects the cost and quality. The goal of the second stage is to minimize the fitness determined by the quality of the candidate circuit $M$
\begin{equation}\label{eq:optproblem}
F(M) = \begin{cases}
			\mathcal{Q}(M), & \text{if $t - \varepsilon \le \mathcal{C}(M) \le t + \varepsilon$}\\
                \infty, & \text{otherwise}
		 \end{cases}
\end{equation}

In order to determine the quality of a given comparison network $M$, we use the quality metric $\mathcal Q(M)$ constructed from the histogram of error distribution $H(M)$ defined above as follows
\begin{equation}
\mathcal Q(M) = \sum_{j = -m+1}^{m-1} j^2 H_{m+j}(M) 
 \label{eq:qc}
\end{equation}
While this function returns zero value if and only if $M$ is the exact median, a positive non-zero value is returned in the other cases. The $j$ value determining the rank weight is squared because it emphasizes the most important part of the histogram and minimizes the worst-case errors.

In contrast to the previous works we targets fully pipelined architecture which needs to consider the number of stages, registers, CAS operations and multiplexors.
Given a candidate comparison network $M$, the implementation cost $\mathcal C(M)$ is determined as follows.
First, the active nodes (CAS) are identified. 
A node is active if at least one of its outputs is connected to (a) the primary output or (b) the input of an active node. 
The ASAP~\cite{Paulin:1989} scheduling algorithm is then used to determine the position of the pipeline registers.
Outputs connected only to inactive nodes are ignored during this step.
The area on a chip required to implement a candidate comparison network can then be determined as $\mathcal C(M) = A_{mx} (2 n_A + n_P) + A_{cmp}(n_A + n_P) + A_{reg} n_R$,
where $n_A$ is the number of active nodes with both outputs connected to either an active node or primary output, $n_P$ is the number of active nodes with one output connected to an inactive node, $n_R$ is the number of required registers and $A_{mx}$, $A_{cmp}$, $A_{reg}$ are the area on chip required to implement a 2-input $w$-bit multiplexer, a magnitude comparator and a register respectively in a chosen target technology.

\section{Results and Discussion}\label{sec:setup}
For statistical evidence, we carried out 20 repetitions of CGP at each design point. 
In total, 800 experimental runs were performed for each median network, each running 30 minutes.
In the case of 9-input and 25-input median, the known optimal implementations consisting of the minimum number of CAS elements were used as the reference initial design point. \footnote{We report results for 9- and 25-input median networks only, the full results are available in \repository.}
The selected networks were implemented in VHDL as fully pipelined streaming architectures operating at 8-bit precision and synthesized to the 45nm technology with 1~GHz clock using Design Compiler.

\subsection{Analysis of Approximate Medians}
The crucial component of the evolutionary search is the calculation of the histogram of error distribution $H(M)$. Although BDDs are generally known for their poor scalability in certain cases, the proposed approach exhibits excellent scalability. 
On average, 400 milliseconds are required to compute the error distribution for $n=49$. 
If we were to use exhaustive simulation instead of the BDDs, we would have to evaluate the response for $2^{49}$ input combinations, which is intractable. The method presented in~\cite{Vasicek2016} requires evaluating $49!$ input combinations, which is more than $10^{62}$ and also intractable.

\begin{figure}[t]
\centering\includegraphics[width=\columnwidth]{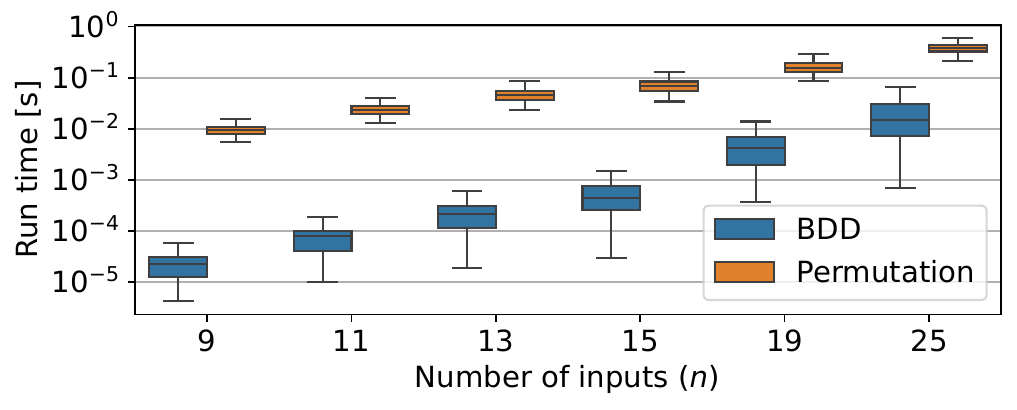}\vspace{-1em}
\caption{Computational complexity of the proposed BDD-based analysis framework compared to the subset of 1000 vectors in permutations \cite{Vasicek2016}.}\label{fig:time}
\end{figure}

Figure~\ref{fig:time} summarizes the run times we measured during our experiments. Each boxplot illustrates the time required to analyze a single candidate solution during the evolutionary search. We compared our proposed exact principle with the permutation testing method presented in \cite{Vasicek2016,vasicek:sekanina:tec}. Even though the permutation testing only examined 1000 random permutations (i.e., 0.2\% and 6$\cdot 10^{-21}$\% of all possible permutations for 9- and 25-input medians, respectively), the BDD-based proposed method is more than two orders of magnitude faster and provides the true $H$ distribution and prove the worst-case errors $d_L, d_H$.

\subsection{Designed Circuits}
\begin{table}[b!]
\setlength{\tabcolsep}{2.5pt}%
\caption{
Parameters of selected Pareto optimal approximations of 
median networks.
The $k$ denotes the number of CAS, $l$ is the latency. The area is given in $\mu\mathrm{m}^2$ and power in mW.}\label{tab:results}
\newcommand{\tabhd}{\toprule 
   &     \multicolumn{4}{c}{Implementation cost} & \multicolumn{4}{c}{Quality indicators}  \\\cmidrule(r){2-5}\cmidrule(l){6-10}
\# & $k$ & $l$ & Area & Pwr & \centering$\mathcal Q$ & $d_L$ & $d_R$ & \centering$h_0$ & $H$ 
 \\ \midrule}
 
\begin{subfigure}[c]{\columnwidth}
\begin{tabularx}{\columnwidth}{c r r R  R  R r r R c}\tabhd 
1 & 19 & 41 & 6272 & 6.44 & 0.00 & 0 & 0 & 1.00 & \hist{9}{0.00,0.00,0.00,0.00,1.00,0.00,0.00,0.00,0.00}\\
2 & 18 & 40 & 5990 & 6.25 & 0.01 & 1 & 1 & 0.98 & \hist{9}{0.00,0.00,0.00,0.01,0.98,0.01,0.00,0.00,0.00}\\
3 & 17 & 37 & 5580 & 5.85 & 0.06 & 1 & 1 & 0.94 & \hist{9}{0.00,0.00,0.00,0.02,0.94,0.04,0.00,0.00,0.00}\\
4 & 16 & 35 & 5227 & 5.43 & 0.10 & 1 & 1 & 0.90 & \hist{9}{0.00,0.00,0.00,0.05,0.90,0.06,0.00,0.00,0.00}\\
5 & 15 & 27 & 4498 & 4.39 & 0.13 & 1 & 1 & 0.87 & \hist{9}{0.00,0.00,0.00,0.06,0.87,0.06,0.00,0.00,0.00}\\
6 & 14 & 23 & 4019 & 3.76 & 0.28 & 1 & 1 & 0.71 & \hist{9}{0.00,0.00,0.00,0.17,0.71,0.12,0.00,0.00,0.00}\\
7 & 12 & 17 & 3203 & 2.86 & 0.56 & 2 & 2 & 0.52 & \hist{9}{0.00,0.00,0.01,0.24,0.52,0.22,0.01,0.00,0.00}\\
8 & 10 & 16 & 2847 & 2.66 & 0.86 & 2 & 2 & 0.39 & \hist{9}{0.00,0.00,0.06,0.28,0.39,0.25,0.02,0.00,0.00}\\
9 & 8 & 9 & 1893 & 1.60 & 1.25 & 2 & 2 & 0.32 & \hist{9}{0.00,0.00,0.07,0.21,0.32,0.29,0.12,0.00,0.00}\\
10 & 6 & 6 & 1413 & 1.14 & 1.95 & 3 & 2 & 0.25 & \hist{9}{0.00,0.00,0.12,0.22,0.25,0.21,0.13,0.06,0.00}\\\midrule
MoM & 12 & 23 & 3760 & 3.64 & 0.43 & 1 & 1 & 0.57 & \hist{9}{0.00,0.00,0.00,0.21,0.57,0.21,0.00,0.00,0.00}\\
\bottomrule
\end{tabularx}
\caption{9-input median circuits}
\end{subfigure}

\begin{subfigure}[c]{\columnwidth}
\begin{tabularx}{\columnwidth}{c r r R r R R r r R c}\tabhd 
1 & 99 & 294 & 38112 & 45.39 & 0.00 & 0 & 0 & 1.00 & \hist{25}{0.00,0.00,0.00,0.00,0.00,0.00,0.00,0.00,0.00,0.00,0.00,0.00,1.00,0.00,0.00,0.00,0.00,0.00,0.00,0.00,0.00,0.00,0.00,0.00,0.00}\\
2 & 89 & 266 & 34101 & 40.49 & 0.05 & 4 & 4 & 0.96 & \hist{25}{0.00,0.00,0.00,0.00,0.00,0.00,0.00,0.00,0.00,0.00,0.00,0.02,0.96,0.02,0.00,0.00,0.00,0.00,0.00,0.00,0.00,0.00,0.00,0.00,0.00}\\
3 & 78 & 347 & 37130 & 49.22 & 0.33 & 5 & 4 & 0.75 & \hist{25}{0.00,0.00,0.00,0.00,0.00,0.00,0.00,0.00,0.00,0.00,0.01,0.13,0.75,0.09,0.01,0.00,0.00,0.00,0.00,0.00,0.00,0.00,0.00,0.00,0.00}\\
4 & 74 & 325 & 34656 & 46.00 & 0.46 & 4 & 5 & 0.66 & \hist{25}{0.00,0.00,0.00,0.00,0.00,0.00,0.00,0.00,0.00,0.00,0.02,0.16,0.66,0.14,0.01,0.00,0.00,0.00,0.00,0.00,0.00,0.00,0.00,0.00,0.00}\\
5 & 67 & 302 & 32020 & 42.80 & 0.64 & 5 & 4 & 0.57 & \hist{25}{0.00,0.00,0.00,0.00,0.00,0.00,0.00,0.00,0.00,0.00,0.03,0.21,0.57,0.16,0.02,0.00,0.00,0.00,0.00,0.00,0.00,0.00,0.00,0.00,0.00}\\
6 & 64 & 224 & 26415 & 32.79 & 0.67 & 5 & 6 & 0.60 & \hist{25}{0.00,0.00,0.00,0.00,0.00,0.00,0.00,0.00,0.00,0.01,0.04,0.20,0.60,0.13,0.01,0.00,0.00,0.00,0.00,0.00,0.00,0.00,0.00,0.00,0.00}\\
7 & 58 & 179 & 22291 & 26.65 & 0.99 & 5 & 5 & 0.48 & \hist{25}{0.00,0.00,0.00,0.00,0.00,0.00,0.00,0.00,0.00,0.01,0.04,0.18,0.48,0.22,0.06,0.01,0.00,0.00,0.00,0.00,0.00,0.00,0.00,0.00,0.00}\\
8 & 50 & 134 & 17779 & 20.34 & 1.42 & 5 & 5 & 0.37 & \hist{25}{0.00,0.00,0.00,0.00,0.00,0.00,0.00,0.00,0.00,0.02,0.07,0.22,0.37,0.22,0.08,0.02,0.00,0.00,0.00,0.00,0.00,0.00,0.00,0.00,0.00}\\
9 & 42 & 90 & 13267 & 13.98 & 2.37 & 6 & 6 & 0.26 & \hist{25}{0.00,0.00,0.00,0.00,0.00,0.00,0.00,0.00,0.01,0.04,0.11,0.21,0.26,0.21,0.10,0.04,0.01,0.00,0.00,0.00,0.00,0.00,0.00,0.00,0.00}\\
10 & 32 & 53 & 8747 & 8.55 & 3.60 & 7 & 7 & 0.20 & \hist{25}{0.00,0.00,0.00,0.00,0.00,0.00,0.00,0.01,0.02,0.07,0.13,0.19,0.20,0.17,0.12,0.06,0.02,0.01,0.00,0.00,0.00,0.00,0.00,0.00,0.00}\\\midrule
MoM & 42 & 83 & 12092 & 12.88 & 1.95 & 4 & 4 & 0.29 & \hist{25}{0.00,0.00,0.00,0.00,0.00,0.00,0.00,0.00,0.00,0.03,0.10,0.22,0.29,0.22,0.10,0.03,0.00,0.00,0.00,0.00,0.00,0.00,0.00,0.00,0.00}\\
\bottomrule
\end{tabularx}
\caption{25-input median circuits}
\end{subfigure}

\end{table}

Parameters of the selected approximate medians are given in Table~\ref{tab:results}.
The reference implementation of the median filter (i.e. the exact implementation) is denoted as \#1 and \textit{median of medians}~\cite{blum} is denoted as MoM. The parameters of software-defined approximate medians~\cite{Vasicek2016} are not included because they are not suitable for the pipelined architecture.
To compare the estimated value with real area, the value of cost function $\mathcal C(M)$ is also given. 
In addition to that, various quality parameters such as the normalized value of quality function $\mathcal Q(M)$, worst-case error ($d_L$, $d_R$), error probability ($h_0$) and error distribution $H(M)$ are included.
The value of the quality indicator $\mathcal Q(M)$ decreases with decreasing number of CAS elements and it seems that the central tendency of the error distribution is adequately captured by the proposed quality metric.

If we reduce the number of CAS by 26\% for 9-input median, i.e. to $k=14$ operations corresponding to the instance \#6 in Table~\ref{tab:results} (a), the output value is correctly determined in more than 72\% of all possible cases.
In the remaining cases, the output value is incorrectly determined as the \xth{4} or \xth{6} lowest element of the list of sorted input elements. Since the median is equal to the \xth{5} lowest element, the distance between the median and the output value is equal to 1 in both cases. 
If we use this approximate median, e.g. in multimedia, the error will hardly be visible in practice.
By using this architecture instead of the exact one described as \#1, we can reduce power consumption by 35\% without compromising output quality.
In some cases, e.g. in the outlier detection task, we can tolerate even more distant errors and use e.g. implementation \#9, which allows us to achieve more than 69\% power improvement.
The MoM \cite{blum} represent widely used approximate implementation of median. These medians for the same hardware cost exhibits lower worst-case error, however, the overall quality is worser because the probability of the exact solution $h_0$ is significantly lower.

Quality in application was evaluated on ten randomly selected test images from the Berkley image dataset~\cite{berkeley:images}. Each image was corrupted by salt-and-pepper noise and random-valued shot noise of 1\%, 5\%, 10\%, 15\% and 20\% intensity. 
The first six instances of the 9-input median have SSIM above 0.97, which is extremely good. In the case of the 25-input medians, we can observe a very small decrease in SSIM. The SSIM remains above 0.9 even for the most approximated instance (instance \#10). This suggests that the noise filtering task could tolerate an even more aggressive approximation.
Note that the 9-input (25-input) MoM exhibits SSIM=0.971 (0.960). This roughly corresponds to the instance \#6 (\#9) with the same implementation cost.

\section{Conclusions}

In this paper, we propose a systematic design method for search-based synthesis of approximate median architectures. The key idea is to reduce the number of CAS elements while introducing a minimum error.
The obtained implementations can be combined with other approaches, e.g. approximate comparators \cite{Monajati:2015} or composition of medians \cite{blum,Adams,Perrot2022}, to further reduce the power consumption.

In order to measure the quality of the approximation, an efficient method for the analysis of the worst-case error and error distribution has been proposed.
Combined with the search-based design approach, our framework is able to produce high-quality approximations with a guaranteed error distribution that is maintained regardless of the chosen bit width. %

\clearpage
\bibliographystyle{IEEEtran}
\bibliography{IEEEabrv,paper}

\begin{thebibliography}{10}
\providecommand{\url}[1]{#1}
\csname url@samestyle\endcsname
\providecommand{\newblock}{\relax}
\providecommand{\bibinfo}[2]{#2}
\providecommand{\BIBentrySTDinterwordspacing}{\spaceskip=0pt\relax}
\providecommand{\BIBentryALTinterwordstretchfactor}{4}
\providecommand{\BIBentryALTinterwordspacing}{\spaceskip=\fontdimen2\font plus
\BIBentryALTinterwordstretchfactor\fontdimen3\font minus \fontdimen4\font\relax}
\providecommand{\BIBforeignlanguage}[2]{{%
\expandafter\ifx\csname l@#1\endcsname\relax
\typeout{** WARNING: IEEEtran.bst: No hyphenation pattern has been}%
\typeout{** loaded for the language `#1'. Using the pattern for}%
\typeout{** the default language instead.}%
\else
\language=\csname l@#1\endcsname
\fi
#2}}
\providecommand{\BIBdecl}{\relax}
\BIBdecl

\bibitem{Ferdowsi:2014}
H.~Ferdowsi, S.~Jagannathan, and M.~Zawodniok, ``An online outlier identification and removal scheme for improving fault detection performance,'' \emph{IEEE Trans. NNLS}, vol.~25, no.~5, pp. 908--919, 2014.

\bibitem{Nikolova:2004:Med}
M.~Nikolova, ``A variational approach to remove outliers and impulse noise,'' \emph{J. Math. Imaging Vis.}, vol.~20, no. 1-2, pp. 99--120, 2004.

\bibitem{Zhang_arxiv.2205.15867}
J.~Zhang, Z.~Su, and L.~Liu, ``Median pixel difference convolutional network for robust face recognition,'' 2022.

\bibitem{Adams}
A.~Adams, ``Fast median filters using separable sorting networks,'' \emph{ACM Trans. Graph.}, vol.~40, no.~4, Jul. 2021.

\bibitem{Mittal:2016}
S.~Mittal, ``A survey of techniques for approximate computing,'' \emph{ACM Comput. Surv.}, vol.~48, no.~4, pp. 62:1--62:33, 2016.

\bibitem{bookaxc}
\BIBentryALTinterwordspacing
\emph{Approximate Circuits: Methodologies and CAD}.\hskip 1em plus 0.5em minus 0.4em\relax Springer International Publishing, 2019. [Online]. Available: \url{http://dx.doi.org/10.1007/978-3-319-99322-5}
\BIBentrySTDinterwordspacing

\bibitem{blum}
\BIBentryALTinterwordspacing
M.~Blum, R.~W. Floyd, V.~Pratt, R.~L. Rivest, and R.~E. Tarjan, ``Time bounds for selection,'' \emph{Journal of Computer and System Sciences}, vol.~7, no.~4, pp. 448--461, 1973. [Online]. Available: \url{https://www.sciencedirect.com/science/article/pii/S0022000073800339}
\BIBentrySTDinterwordspacing

\bibitem{Perrot2022}
G.~Perrot, S.~Domas, and R.~Couturier, ``How separable median filters can get better results than full 2d versions,'' \emph{The Journal of Supercomputing}, vol.~78, no.~7, pp. 10\,118--10\,148, Jan. 2022.

\bibitem{salvador18}
G.~Salvador, J.~M. Chau, J.~Quesada, and C.~Carranza, ``Efficient gpu-based implementation of the median filter based on a multi-pixel-per-thread framework,'' in \emph{2018 IEEE Southwest Symposium on Image Analysis and Interpretation (SSIAI)}, 2018, pp. 121--124.

\bibitem{Monajati:2015}
M.~Monajati, S.~M. Fakhraie, and E.~Kabir, ``Approximate arithmetic for low-power image median filtering,'' \emph{Circuits Syst. Signal Process.}, vol.~34, no.~10, pp. 3191--3219, 2015.

\bibitem{vasicek:sekanina:tec}
Z.~Vasicek and L.~Sekanina, ``Evolutionary approach to approximate digital circuits design,'' \emph{IEEE Trans. on Evolutionary Computation}, vol.~19, no.~3, pp. 432--444, 2015.

\bibitem{Vasicek2016}
Z.~Vasicek and V.~Mrazek, ``Trading between quality and non-functional properties of median filter in embedded systems,'' \emph{Genetic Programming and Evolvable Machines}, vol.~18, no.~1, pp. 45--82, Jul. 2016.

\bibitem{Knuth:ACP}
D.~E. Knuth, \emph{The Art of Computer Programming, Volume 3: (2Nd Ed.) Sorting and Searching}.\hskip 1em plus 0.5em minus 0.4em\relax Addison Wesley Longman Publishing Co., Inc., 1998.

\bibitem{aproxcomp22}
A.~Bosio, D.~M{\'{e}}nard, and O.~Sentieys, Eds., \emph{Approximate Computing Techniques}.\hskip 1em plus 0.5em minus 0.4em\relax Springer International Publishing, 2022.

\bibitem{Sekanina2021}
L.~Sekanina, ``Evolutionary algorithms in approximate computing: A survey,'' \emph{Journal of Integrated Circuits and Systems}, vol.~16, no.~2, pp. 1--12, Aug. 2021.

\bibitem{miller:cgp:book}
J.~F. Miller, \emph{Cartesian Genetic Programming}.\hskip 1em plus 0.5em minus 0.4em\relax Springer-Verlag, 2011.

\bibitem{Paulin:1989}
P.~G. Paulin and J.~P. Knight, ``Force-directed scheduling for the behavioral synthesis of asics,'' \emph{IEEE Trans. on Computer-Aided Design of Integrated Circuits and Systems}, vol.~8, no.~6, pp. 661--679, 1989.

\bibitem{berkeley:images}
D.~Martin, C.~Fowlkes, D.~Tal, and J.~Malik, ``A database of human segmented natural images and its application to evaluating segmentation algorithms and measuring ecological statistics,'' in \emph{Proc. 8th Int'l Conf. Computer Vision}, vol.~2, July 2001, pp. 416--423.

\end{thebibliography}

\end{document}